\documentclass[aps,a4paper]{revtex4-2}
\usepackage{epsfig}
\usepackage{graphicx}
\usepackage{amsmath,amssymb,color}
\usepackage[english]{babel}

\parskip=\medskipamount

\newcommand{\eq}[1]{(\ref{#1})}
\newcommand{\fig}[1]{Fig.~\ref{#1}}

\newcommand{\be}{\begin{equation}}
\newcommand{\ee}{\end{equation}}

\newcommand\disp{\displaystyle}

\newcommand{\la}{\left<}
\newcommand{\ra}{\right>}
\newcommand{\eps}{\varepsilon}

\newcommand{\re}{\textrm{Re}\,}
\newcommand{\im}{\textrm{Im}\,}

\begin{document}

\title{\textsf{\small On the basis of the talk delivered by one of authors (S.N.) at the workshop ``Statistical mechanics and its applications'', Dilijan, Armenia (October 2022)} \\
\noindent\rule{12cm}{0.4pt}
\bigskip \\ 
BKT in Phyllotaxis}

\author{A. Flack and S. Nechaev}

\affiliation{LPTMS, CNRS--Universit\'e Paris-Saclay, 91405 Orsay Cedex, France}

\date{\today}

\begin{abstract}

We discuss a two-parameter renormalization group (RG) consideration of a phyllotaxis model in the framework of the ``energetic approach'' proposed by L. Levitov in \cite{levitov}. Following L. Levitov, we consider an equilibrium distribution of strongly repulsive particles on the surface of a finite cylinder and study the redistribution of these particles when the cylinder is squeezed along its axis. We construct explicitly the $\beta$-function of a given system in terms of the modular Dedekind $\eta$-function. On basis of this $\beta$-function we derive the equations describing the RG flow in the vicinity of the bifurcation points between different lattices. Analyzing the structure of RG equations, we claim emergence of Berezinskii-Kosterlitz-Thouless (BKT) transitions at strong compression of the cylinder. 

\end{abstract}

\maketitle


\section{Introduction}
\label{sect:01}

Amazing connection of cell packing with Fibonacci sequences, known as phyllotaxis \cite{phyllotaxis} was observed a long time ago in the works of naturalists and remains till now one of the most known manifestations of number theory in natural science. The generic description of growing plants based on symmetry arguments allowed researchers to uncover the role of Farey sequences in the plant's structure formation (see, for example, \cite{phyllo2, phyllo3, phyllo4}), however, the question why the nature selects the Fibonacci sequence, among other possible Farey ones, was hidden until modern time. A tantalizing answer to this question has been given by L. Levitov in 1990 in \cite{levitov}, who proposed an ``energetic'' approach to the phyllotaxis, suggesting that the development of a plant is connected with an effective motion along the optimal path on a Riemann surface associated with the energetic relief of growing tissue.

The energetic mechanism suggested in \cite{levitov1} was applied later in \cite{levitov3} to the investigation of the geometry of flux lattices pinned by layered superconductors. It has been shown that under the variation of a magnetic field, the structure of the flux lattice can undergo a sequence of rearrangements encoded by the Farey numbers. However, lattices emerging in sequential rearrangements are characterized by the specific subsequence of the Farey set, namely, by the Fibonacci numbers. Very illuminating experiments have been provided in \cite{rotating} for lattice formed by drops in rotating liquid, and in \cite{cactus} for the equilibrium structure of a ``magnetic cactus''.

The general classification of RG flows rhymes with the development of bifurcations (``catastrophes'') over time in the theory of dynamical systems -- see, for instance \cite{gukov2017rg}. In the catastrophe theory there are focuses, saddles, limits cycles and other attributes of the singularity theory, with corresponding fixed points, RG cycles and more exotic RG behavior. For instance, recently the RG counterparts of homoclinic orbits in the theory of dynamical systems have been found in the field theory \cite{jepsen2021homoclinic}, they also provide examples of chaotic RG flows \cite{bosschaert2022chaotic}. The phenomenon of incommensurability is also known in the theory of dynamical systems. Following the same logic, one could expect the existence of RG counterpart of the incommensurability. Indeed, the RG approach was successful in describing the incommensurable patterns  in a Harper equation for the electron in a crystal in presence of a magnetic field \cite{wilkinson1984critical, wilkinson1987exact} where it was argued that the tunneling in the phase space is the crucial ingredient. 

In an overwhelming majority of physical systems \cite{aubry1983devil, bak1982commensurate} the inmommesurability manifests itself in an emergence of a ``Devil's staircase''. The geometric signature of the incommensurability is the Riemann-Thomae (RT) function which emerges in spectra of sparse systems of various physical origin.  Meanwhile, the Riemann-Thomae function also appears as a probability distribution in a plethora of fundamental problems, such as stability diagram in fractional quantum Hall effect \cite{Tao1, Tao2}, interactions of non-relativistic ideal anyons with rational statistics in the ``magnetic gauge'' approach \cite{lundholm}, quantum $1/f$ noise and Frenel-Landau shift \cite{planat}, distribution of quotients of reads in DNA sequencing experiment \cite{dna}, frequency of specific subgraphs counting in the protein-protein network of a \textit{Drosophilla} \cite{drosophilla}. Though the degree of similarity with the original RT function could vary, and experimental profiles may drastically depend on the peculiarities of each particular physical system, a general probabilistic scheme resulting in emergence of the fractal hierarchical distribution can be considered as the manifestation of number-theoretic laws in nature.

Often two real parameters of a 2D RG flow are combined into the single complex parameter, $\tau$, which can be interpreted as the modulus of the complex structure for an auxiliary elliptic curve. The familiar examples are: the Anderson localization problem with the time symmetry breaking (TSB) term \cite{Altland2015topology}, the integer quantum Hall effect (IQHE) \cite{pruisken1984localization, levine1984theory}, and the Yang-Mills theory with the TSB $\theta$-term \cite{montonen1977magnetic, cardy1982phase}. In all these examples the real part of the complex parameter is the TSB parameter. We suggest a bit more general perspective and propose to consider the following generic complex (modular) parameter:
\be
z = {\rm [topological ~ term]} + i\,{\rm [disorder]},
\label{eq:tau}
\ee
hence we unite the topology and the disorder in the RG flow. At any fixed value $z=x+i y$ the partition functions of considered systems fully enjoy symmetries of the $SL(2,Z)$ modular group and hence are the modular functions. However when $x$ and/or $y$ run over time and depend on some scale, $\mu$, the situation is more subtle. It general, the RG flow involves two $\beta$-functions and is described by the set of equations
\be
\begin{cases}
\disp \frac{dx}{d \ln\mu}= \beta_{x}(x,y) \medskip \\
\disp \frac{dy}{d \ln\mu}= \beta_{y}(x,y)
\label{eq:beta}
\end{cases}
\ee
Typically, the disorder term enjoys both the perturbative and non-perturbative remormalizations, while the topological parameter is renormalized only non-perturbatively. There are some known patterns of $\beta$-functions with such properties: 
\begin{itemize}
\item For the Berezinskii-Kosterlitz-Thouless (BKT) transition one has: 
\be
\begin{cases}
\disp \beta_u= -c_1 uv \\
\disp \beta_v=-c_2 u^2
\label{eq:beta2}
\end{cases}
\ee
\item For the ``Russian Doll'' model which is the toy example of the system with the cyclic RG flows (see, \cite{bulycheva2014limit} for review), the RG flow is discrete
\be
\begin{cases}
g_{N+1}= g_N + \frac{1}{N}(g_N^2 +\theta_N^2) \\
\theta_{N+1}=\theta_N
\label{eq:beta1}
\end{cases}
\ee
\end{itemize}

We focus our attention on specific limit of RG flows when the non-perturbative renormalization coming from instanton-like contributions dominates -- see, for example, \cite{wilkinson1987exact}. This happens in all examples when $y=\im{\tau}\to 0$ which means that we are looking at the limit of a weak disorder in some frame, and the modular parameter is mainly governed by the ``winding-like'' term. The details are model-dependent, however in all cases  this term has one and the same physical meaning: it serves for counting topological defects.  

One possible pattern behind the Riemann-Thomae function and the Devil's staircase is as follows. Consider a physical problem, for example the fractional quantum Hall effect (FQHE), and push the system into the particular limit in the parameter space. For FQHE this is the so-called ``thin torus limit'' -- see for example \cite{Tao2}. The system hosts some defects, and in the limit under consideration defects form a lattice which is a Wigner crystal in the thin torus limit of FQHE. Consider now the propagation of a probe particle through the sample which can be investigated, for instance, by analyzing its spectral density. The modular $SL(2,Z)$ group acts in the parameter space of this system. The imaginary part $\im \tau$ of the modular parameter $\tau$ gets identified with some function of disorder, while the real part $\re \tau$ corresponds to the chemical potential for the topological charge relevant for the studied problem. The motion of the probe particle in the crystal of defects can be mapped onto the motion in the fundamental domain of $SL(2,Z)$ and the rearrangements of the lattice can be treated by analyzing the RG flow in the vicinity of transition points which are identified with points of lattice bifurcations. Generally speaking, from the probe particle perspective, the rearrangement of the lattice can be studied by varying the chemical potential of defects (or of their number).

The paper is organized as follows: in Section \ref{sect:02} we define the model, discuss the modular properties of the potential acting between particles, construct the $\beta$-function and show explicitly its scale-invariant structure; in Section \ref{sect:03} we derive the RG flow equations in the vicinity of critical (saddle) points and demonstrate the convergence of these equations at small $y$ to the ones describing the BKT transition in $XY$ model; finally, in Section \ref{sect:04} we summarize the obtained results and in addition provide arguments showing that the variable $y$ besides its geometrical sense, can be related to the disorder in a simple spectral problem. 

\section{Phyllotaxis, modular invariance and $\beta$-function}
\label{sect:02}

\subsection{The model}
\label{sect:02a}

\begin{figure}[ht]
\centerline{\includegraphics[width=14cm]{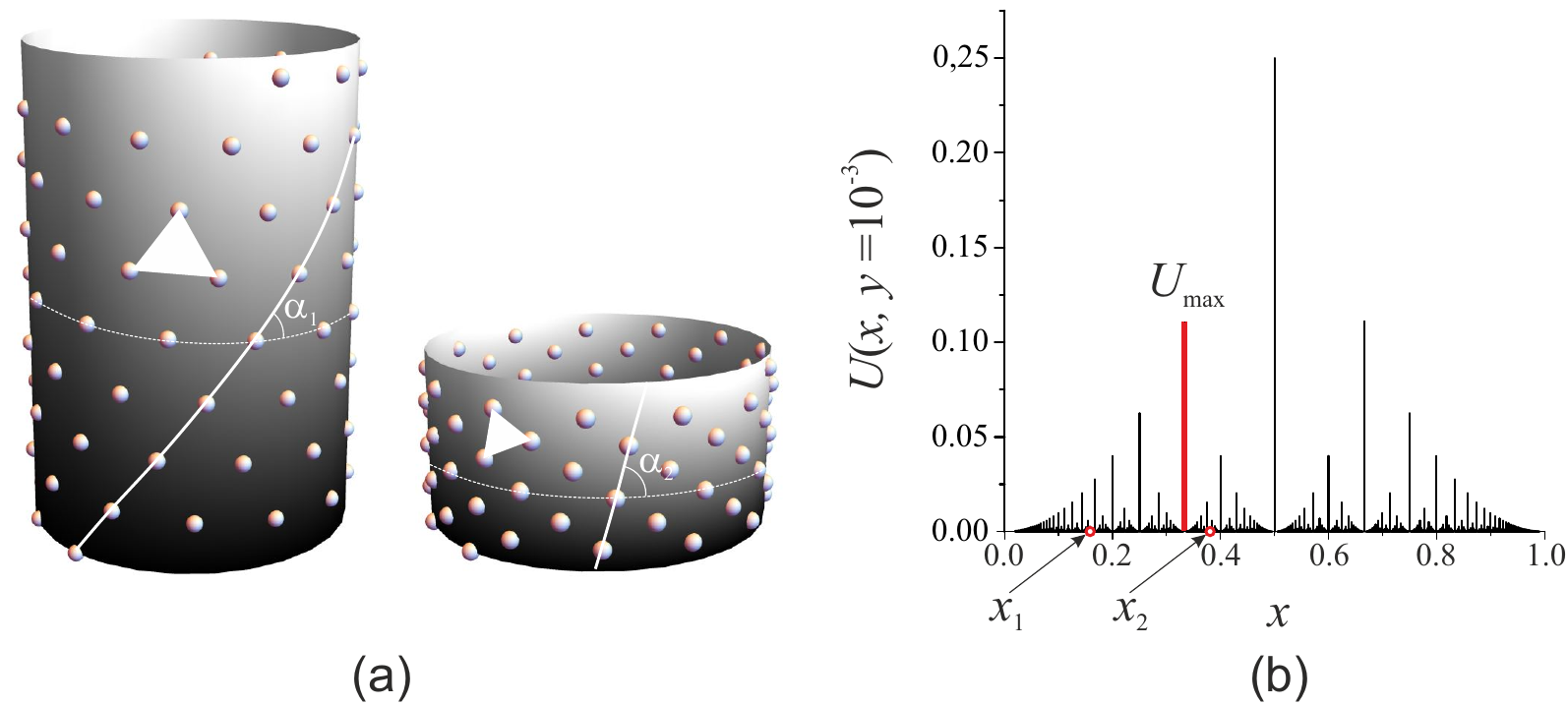}}
\caption{(a) Repulsive points distributed on the surface of the cylinder form a lattice, characterized by the parameter $\alpha$, with a minimal energy. The lattice is rearranged when the cylinder is compressed along its vertical axis; (b) Dependence $U(x, y={\rm const})$ defined in \eq{eq:08} for the compressed lattice ($y=10^{-3} \ll 1$ and $\beta=1$) as a function of $x$.}
\label{fig:01}
\end{figure}

We consider, following L. Levitov \cite{levitov1}, the model system of $N$ strongly repulsive particles disposed and equilibrated at the surface of a cylinder of fixed diameter, $D$, and height, $H$ and look at the rearrangement of these particles when the cylinder is compressed along its height under the condition that $N$ and $D$ remain unchanged -- see \fig{fig:01}a. This model can be regarded as a kind of modification of a famous Tammes problem dealt with a packing a given number of points on the surface of a sphere such that the shortest distance between points is maximized. The Tammes problem is known in plant's geometry  since 1930 \cite{tammes} and it itself can be viewed as a particular case of the generalized Thomson problem \cite{thomson} of minimizing the total Coulomb energy of charged particles distributed on the surface of a sphere. The advantage and novelty of L. Levitov's phyllotaxis model with respect to the Thomson-Tammes system is two-fold: (a) cylindrical lattice is described by two parameters ($D$ and $H$) and one can change them independently, and (b) the equilibrium lattices on the cylinder are transformed under the action of the group $SL(2,Z)$, which essentially simplifies the consideration of rearrangement of lattices when changing $D$ and/or $H$.

At the continuous compression of the cylinder, for each height, particles form a triangular ``Abrikosov'' lattice with minimal energy \cite{abrikosov}. Different lattice topologies correspond to local minima of the system's energy $U(x,y)$ and are parametrized by the modular parameter, $z=x+iy$, where $x=f_1(D,H)$,  $y=f_2(H)$  and $f_1$ and $f_2$ are some functions to be defined below. The minima of the potential $U(z)$ are clustered in nested basins: larger basins consist of smaller basins, each of these consists of even smaller ones, etc. The basins of local energy minima are separated by a hierarchically arranged set of barriers: large basins are separated by high barriers, and smaller basins within each larger one are separated by lower barriers. The geometry which fixes taxonomic (i.e. hierarchical) tree-like relationships between elements is called ``ultrametric''  \cite{ultra1}.

Here we provide an explicit construction of the energetic relief in a phase space of all possible patterns of compressed lattices and demonstrate that the ground state is related to the deepest valley in $\Gamma$. The lattice rearrangement caused by the cylinder compression along its axis is associated with the RG flow on the manifold $\Gamma$, which shares the modular properties. At each height, $H$, particles on the cylindrical surface form a lattice with a minimal energy. For strongly squeezed lattices ($y\to 0$) the corresponding energy, $U(x,y)$, has a sharp maximum corresponding to the barrier at every rational point, $x = \frac{m}{n}$ as it is shown in \fig{fig:01}b. One sees that $U(x,y)$ demonstrates the hierarchical behavior, which should be understood as follows: the transitions between two arbitrary local minima at $x_1$ and $x_2$, are determined by the passage over the highest barrier $U_{\rm max}(x_1,x_2)$, separating the points $x_1$ and $x_2$.

The outline of our upcoming discussion is as follows.  We begin with the derivation of the potential $U(x,y)$ separating valleys between different equilibrium configurations of particles on the cylinder. Taking into account the invariance of the potential $U(\tau)$ under the action of the group $SL(2,Z)$, we show that $U(\tau)$ plays a role of a $\beta$-function of the system. The explicit form of the $\beta$-function allows us to derive the RG equations in the vicinity of saddle points of the potential $U(\tau)$ and solve these equations in a general form at any $x$ and $y$. Analyzing the structure of obtained RG equations we demonstrate that they tend to the RG equations of $XY$-model describing the BKT transition in a strong compression limit (i.e. when $y\to 0$).

\subsection{The potential}
\label{sect:02b}

Any particle on the cylinder can be parameterized by a pair $(h_n,\alpha_n\, \{\mathrm{mod}\, 2\pi\})$, where $n\in \mathbb{N}$, subject that all particles are arranged according to monotonic growth of the height, $h_n$. Projecting the cylindrical surface conformally onto the plane, we get new coordinates, $\mathbf{r}_{n,m}(x,y)$, of particles on the planar lattice,
\be
\mathbf{r}_{n,m}(x,y) = \left(\frac{m + n x}{\sqrt{y}},\, n\sqrt{y} \right), \quad \{m,n\} \in \mathbb{Z}^2
\label{eq:01}
\ee
where the connection between cylindrical and planar lattices is set by the following change of variables:
\be
x=\frac{\alpha}{2\pi}, \quad y=\frac{h}{2\pi} \quad (y>0)
\label{eq:02}
\ee
Strong repulsive potential acting between particles can be approximated by the conformally-invariant $1/r^2$ potential. Consider two arbitrary particles one of which is located at the origin of the $(x,y)$-plane and the second -- at some point $(x_{m,n}, y_{m,n})$. Suppose that the potential $U({\bf r}_{m,n})$ has the following form:
\be
U({\bf r}_{m,n}) = \frac{q}{{\bf r}^2_{m,n}}
\label{eq:03}
\ee
where $q>0$ is some arbitrary parameter having sense of a charge. The energy of a whole lattice reads
\be
U(x,y) = \sum_{\{m,n\} \in \mathbb{Z}^2 \backslash \{0, 0\}} U(x_{m,n},y_{m,n}) = \sum_{\{m,n\} \in \mathbb{Z}^2 \backslash \{0, 0\}} \frac{q}{{\bf r}^2_{m,n}}
\label{eq:03a}
\ee
Substituting \eq{eq:01} into \eq{eq:03}, we get:
\be
U(x,y) = \sum_{\{m,n\} \in \mathbb{Z}^2 \backslash \{0, 0\}} \frac{qy}{(m+n x)^2 + y^2 n^2} 
\label{eq:04}
\ee

Recall now the definition of the non-holomorphic Eisenstein series, $E(z,s)$, \cite{eisen}:
\be
E(z,s) = \sum_{\{m,n\} \in \mathbb{Z}^2 \backslash \{0, 0\}} \frac{y^s}{|n z + m|^{2s}};
\qquad z = x+iy
\label{eq:05}
\ee
where $E(z,s)$ is a function of $z=x+iy$ and is defined in the upper half-plane $y>0$ for all $\re(s)>1$.  

The non-holomorphic Eisenstein series of weight 0 and level 1 can be analytically continued to the whole complex $s$-plane with one simple pole at $s=1$. Notably $E(z,s)$, as function of $z$, is the $SL(2,\mathbb{Z})$--automorphic solution of the hyperbolic Laplace equation:
\be
-y^2 \left(\frac{\partial^2}{\partial x^2}+\frac{\partial^2}{\partial y^2}\right) E(x,y, s) = s(1-s)\; E(x,y, s)
\label{eq:06}
\ee
The residue of $E(z, s)$ at $s=1$ is known as the first Kronecker limit formula \cite{epstein, siegel, motohashi}. Explicitly, it reads at $s\to 1$:
\be
E(z, s\to 1) = \frac{\pi}{s-1} + 2\pi\left(\gamma + \ln 2 - \ln \left(y^{1/2}|\eta(z)|^2\right)\right) + O(s-1)
\label{eq:07}
\ee
where $\gamma$ is the Euler constant and $\eta(z)$ is the Dedekind $\eta$-function. Equation \eq{eq:07} establishes the important connection between the Eisenstein series and the Dedekind $\eta$-function, which we exploit below. Namely, comparing \eq{eq:04} to \eq{eq:05}, we conclude that
\be
U(x,y) \approx q E(x+iy, s\to 1) \to 4\pi q \ln \left(y^{1/4}|\eta(x+iy)|\right)
\label{eq:08}
\ee

Let us remind that the Dedekind $\eta$-function is defined as follows:
\be
\eta(z)=e^{\pi i z/12}\prod_{n=0}^{\infty}(1-e^{2\pi i n z})
\label{eq:09}
\ee
The argument $z=x+iy$ is called the modular parameter, and $\eta(z)$ is defined for all $y>0$. The function $\eta(z)$ is invariant with respect to the action of the modular group $SL(2,\mathbb{Z})$:
\be
\begin{cases}
\disp \eta (z+1)=e^{\pi i z/12}\;\eta(z) \medskip \\
\disp \eta\left(-\frac{1}{z}\right) = \sqrt{-i}\; \eta(z)
\end{cases}
\label{eq:10}
\ee
In general,
\be
\eta\left(\frac{az+b}{cz+d}\right) = \omega(a,b,c,d)\; \sqrt{cz + d}\; \eta(z)
\label{eq:11}
\ee
where $ad-bc=1$ and $\omega(a,b,c,d)$ is a 24th degree root of unity \cite{dedekind}.

\subsection{The $\beta$-function}
\label{sect:02c}

The construction of the $\beta$-function implies finding the function which is invariant with respect to RG transformations. The natural candidate for the $\beta$-function is the potential $U(x,y)$. To see the self-similarity of $U(x,y)$ along the RG flow (i.e. at $y$ changing from $+\infty$ down to 0), we consider the function $U(x,y|{\cal D}_1)$ in some initial domain ${\cal D}_1$ and compare it with its own part $U(x,y|{\cal D}_2)$ in a smaller domain ${\cal D}_2$. It is always possible to find a conformal transform $[(x,y)\in {\cal D}_2] \to [(x',y')\in{\cal D}_1]$ constructed on the basis of generators of $SL(2,Z)$ such that $U(x', y'|{\cal D}_2)\to U(x,y|{\cal D}_2)$. Below we demonstrate the corresponding construction on a particular example.

\begin{figure}[ht]
\centerline{\includegraphics[width=15cm]{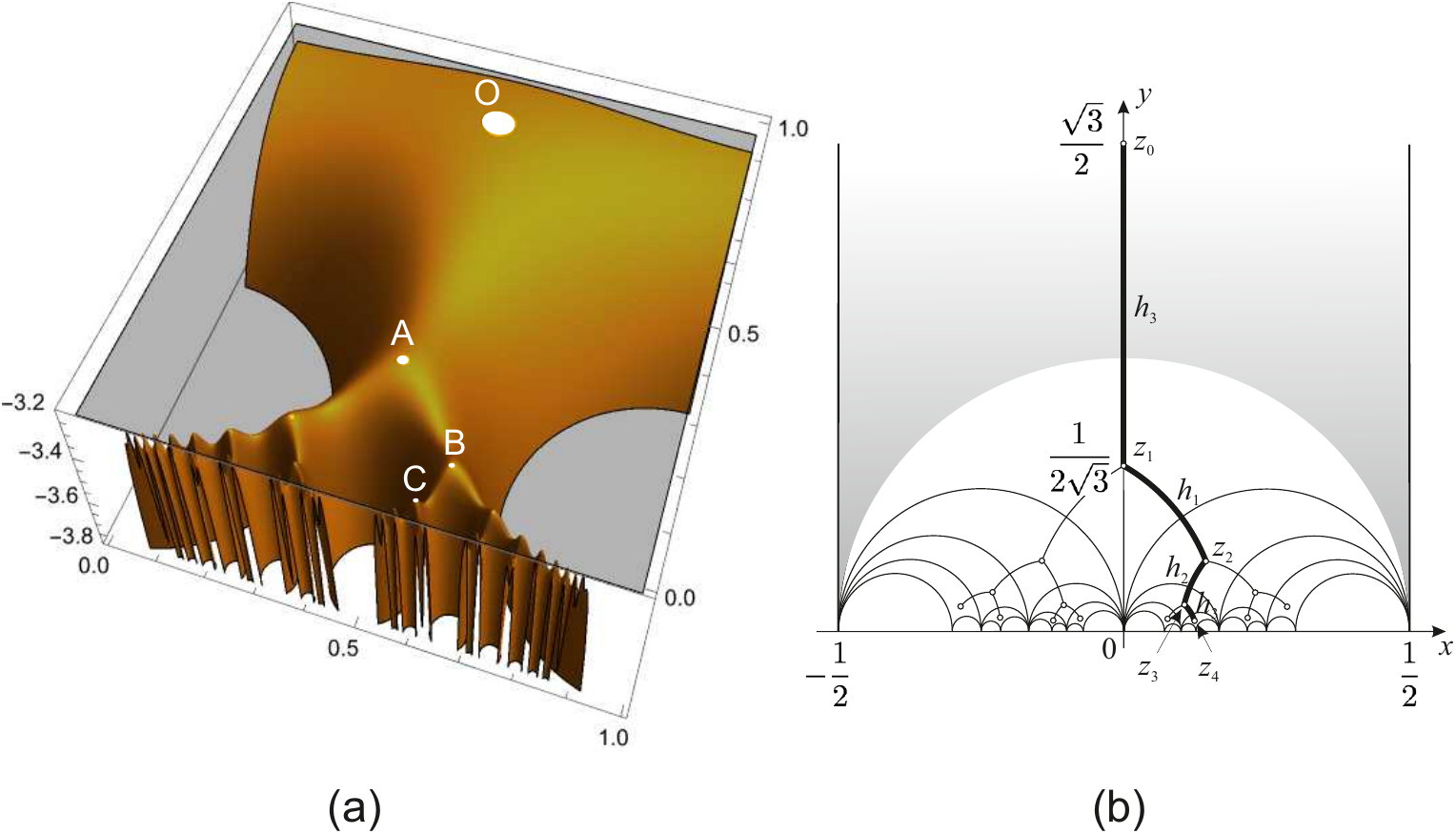}}
\caption{(a) 3D view of the function $U(x,y|{\cal D}_1)$ in the domain ${\cal D}_1=(0<x<1, 0<y<1)$. For better visualization only the part of the function $U(x,y|{\cal D}_1)$ lying in the interval $U(x,y|{\cal D}_1)>-3.8$ is plotted.}
\label{fig:02}
\end{figure}

\begin{figure}[ht]
\centerline{\includegraphics[width=15cm]{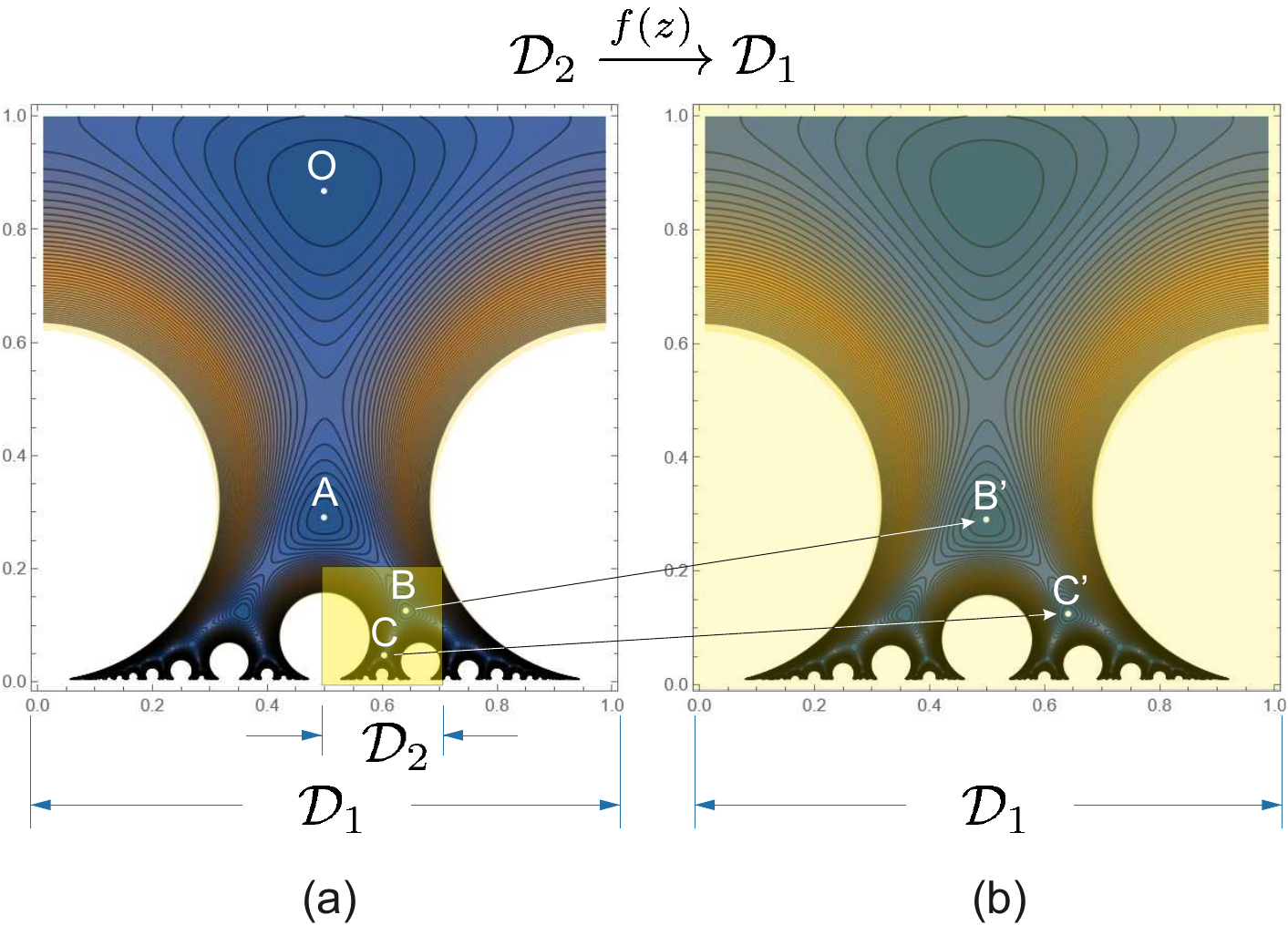}}
\caption{(a) Contour plot of the function $U(x,y|{\cal D}_1)$ in the domain ${\cal D}_1=(0<x<1, 0<y<1)$. The function $U(x,y|{\cal D}_2)$ in the domain ${\cal D}_2$ shown by yellow square in the panel (a) is conformally mapped by the function $f(z)$ (see \eq{eq:12}) onto $U(x\,y|{\cal D}_1)$ in the domain ${\cal D}_1$ as it is shown in the panel (b). The corresponding mapping demonstrates the scale invariance of $U(x,y)$.}
\label{fig:03}
\end{figure}

In \fig{fig:02}a we provide the generic 3D view of the function $U(x,y|{\cal D}_1)$ (at $q=1$) in the domain ${\cal D}_1=(0<x<1, 10^{-2}<y<1)$. All local maxima, $(x_j,y_j)$ of the relief $U(x,y)$ depicted by white points, lie at the level $U(x_j,y_j)\approx -3.248$. For better visualization only the part of the function $U(x,y)$ bounded from below, namely $U(x,y)>-3.8$ is drawn in \fig{fig:02} and in \fig{fig:03} in all panels. Coordinates of particular local maxima shown in \fig{fig:02} are: $O\left(\tfrac{1}{2},\tfrac{\sqrt{3}}{2}\right)$, $A\left(\tfrac{1}{2},\tfrac{1}{2\sqrt{3}}\right)$, $B\left(\tfrac{9}{14},\tfrac{\sqrt{3}}{14}\right)$, $C\left(\tfrac{23}{38},\tfrac{\sqrt{3}}{38}\right)$. In figure \fig{fig:02}b we depict a number of fundamental domains of the group $SL(2,Z)$ together with exact locations of their centers (points $z_0,z_1,z_2,z_3, z_4$) which exactly match the local maxima of the relief $U(x,y)$ in \fig{fig:02}a. We will return to the determination of coordinates of these points at the end of this subsection.

The contour plot of the function $U(x,y|{\cal D}_1)$ in the domain ${\cal D}_1$ is shown in \fig{fig:03}a. To demonstrate the scale invariance (the self-similarity) of $U(x,y)\equiv U(z)$, where $z=x+iy$, we select a new (smaller) region, ${\cal D}_2$, designated by the yellow square in \fig{fig:03}a and seek for a conformal transform $z'=f(z)$ which maps $U(z|{\cal D}_2)$ onto $U(z'|{\cal D}_1)$ as it is shown in \fig{fig:03}a,b. 

For domains ${\cal D}_1=(0<x<1, 10^{-2}<y<1)$ and ${\cal D}_2= \left(x_C - 0.1<x<x_C+0.1,\, 10^{-3}<y<0.2\right)$, where $x_C=\tfrac{23}{38}$ is the $x$-coordinate of the point $C\left(\tfrac{23}{38},\tfrac{\sqrt{3}}{38}\right)$,  the mapping ${\cal D}_2\to {\cal D}_1$ is realized via the conformal transform
\be
z'=f(z)=\frac{z-1}{4z-3}
\label{eq:12}
\ee
The contour plot of the function $U(x',y')$ defined by \eq{eq:08} is presented in \fig{fig:03}b. The variables $(x',y')$ are defined by the folowing equations: 
\be
\begin{cases}
\disp x'=\re \frac{z-1}{4z-3}=\frac{3 - 7 x + 4 x^2 + 4 y^2}{9 - 24 x + 16 x^2 + 16 y^2} \medskip \\ 
\disp y' = \im \frac{z-1}{4z-3}=\frac{y}{9 - 24 x + 16 x^2 + 16 y^2}
\end{cases}
\label{eq:13}
\ee
and the variables $(x,y)$ lie in the domain ${\cal D}_2=\left(\frac{23}{38} - 0.1<x<\frac{23}{38}+0.1,\, 10^{-3}<y<0.2\right)$. As one sees, the plot in \fig{fig:03}b exactly coincides with the one in \fig{fig:03}a in the whole domain ${\cal D}_1=(0<x<1, 0<y<1)$.  Substituting $B\left(x=\tfrac{9}{14},y=\tfrac{\sqrt{3}}{14}\right)$ into \eq{eq:13} we get $B'\left(x'=\tfrac{1}{2},y'=\tfrac{1}{2\sqrt{3}}\right)$ and for 
$C\left(x=\tfrac{23}{38},y=\tfrac{\sqrt{3}}{38}\right)$ we get $C'\left(x'=\tfrac{9}{14},y'=\tfrac{\sqrt{3}}{14}\right)$ -- see \fig{fig:03}. So, we can conclude that the domain ${\cal D}_2$ is expanded onto the domain ${\cal D}_1$ such that the structure of the potential $U(x,y)$ remains completely scale-invariant.

Generically all local maxima (points $z_1,z_2,...$ in \fig{fig:02}b) can be constructed via successive reflections of the fundamental domain of the free group $\Gamma_2$ as it is shown in \fig{fig:02}b. The corresponding Cayley graph is a 3-branching Cayley tree. Recall that the 3-branching Cayley tree can also be viewed as the Cayley graph of the group $\Lambda$, which has the free product structure: $\Lambda \sim \mathbb{Z}_2 \otimes \mathbb{Z}_2 \otimes \mathbb{Z}_2$, where $\mathbb{Z}_2$ is the cyclic group of second order. The matrix representation of generators $h_1, h_2, h_3$ of the group $\Lambda$ is well known: 
\be 
h_1 = \left(\begin{array}{cc} 1 & 1 \medskip \\ 0 & -1 \end{array}\right); \qquad 
h_2 = \left(\begin{array}{cc} 1 & -1 \medskip \\ 0 & -1 \end{array}\right); \qquad 
h_3 = \left(\begin{array}{cc} 0 & \tfrac{1}{2} \medskip \\ 2 & 0 \end{array}\right)
\label{eq:14}
\ee
Taking the point, $z_0=\tfrac{\sqrt{3}}{2}\,i$, we can find  its image, $z_N$, after $N$ recursive applications of generators from the set $\{h_1,h_2,h_3\}$ according to the following formula:
\be
z_N=\frac{1}{2}+\begin{cases} \disp \frac{a_N  \bar{z}_0 + b_N}{c_N\bar{z}_0 + d_N} & \mbox{for $N = 2k-1$, $k=1,2,...$} \medskip \\ 
\disp \frac{a_N  z_0 + b_N}{c_N z_0 + d_N} & \mbox{for $N = 2k$, $k=1,2,...$} 
\end{cases}
\label{eq:15}
\ee
where $\bar{z}$ means complex conjugation of $z$ and $\{a_N, b_N, c_N, d_N\}$ are the coefficients of the matrix 
\be
\left(\begin{array}{cc} a_N & b_N \\ c_N & d_N \end{array} \right) = \overbrace{h_3 h_2 h_1 h_3...}^{N~{\rm terms}} 
\label{eq:16}
\ee
Using \eq{eq:14}-\eq{eq:16} we reproduce the coordinates of the points $A,B,C$ in \fig{fig:02}. The sequence which converges to the Golden ratio is as follows:
\be
\left(\begin{array}{cc} a_{3M} & b_{3M} \\ c_{3M} & d_{3M} \end{array} \right) = \overbrace{\left(h_3 h_2 h_1\right)\left(h_3 h_2 h_1\right) ... \left(h_3 h_2 h_1\right)}^{N~{\rm terms}} = \left(h_3 h_2 h_1\right)^{3M}
\label{eq:17}
\ee
where $N=3M$, $M=1,2,3,...$.  The limiting value of $x_{\infty}=\re z_{N\to \infty}$ is the Golden ratio:
\be
z_{\infty} = \frac{1}{2} + \frac{1}{2}\lim_{M\to\infty} \frac{a_{3M} c_{3M}+ b_{3M} d_{3M}}{c_{3M}^2+d_{3M}^2} = \frac{1}{2}\left(\sqrt{5}-1\right) \approx 0.618034...
\label{eq:18}
\ee
The sequence of ``zigzag'' reflections  is encoded in the continued fraction expansion of the Golden ratio, $\phi$:
\be 
\phi=\frac{1}{2}(\sqrt{5}-1)=\cfrac{1}{1+\cfrac{1}{1+\cfrac{1}{1+\cfrac{1}{1+\cdots}}}}
\label{eq:19}
\ee
where interlacing odd and even ``1'' correspond to the left and right turns of a zigzag path in \fig{fig:02}b.

\section{RG equations and a signature of BKT transitions}
\label{sect:03}

Understanding RG flow as adiabatic particle's dynamics (APD) in a complex potential is very useful in studying the behavior of RG flows in the vicinity of critical points which can be regarded as bifurcation points in the APD problem. Here we derive the corresponding RG equation for the potential $U(x,y) = 4\pi q \ln (y^{1/4} |\eta(x+iy)|)$ defined in \eq{eq:08}. The function $U(z)$, where $z=x+iy$, plays the role of a $\beta$-function which remains invariant under the action of the group $SL(2,Z)$ in particular when $y$ tends to 0. Recall, that in the phyllotaxis problem changing $y$ from $=+\infty$ down to 0 can be interpreted as the re-distribution of the system of repulsive particles (equilibrated at the surface of the cylinder) when the cylinder is squeezed along its axis. 

\begin{figure}[ht]
\centerline{\includegraphics[width=16cm]{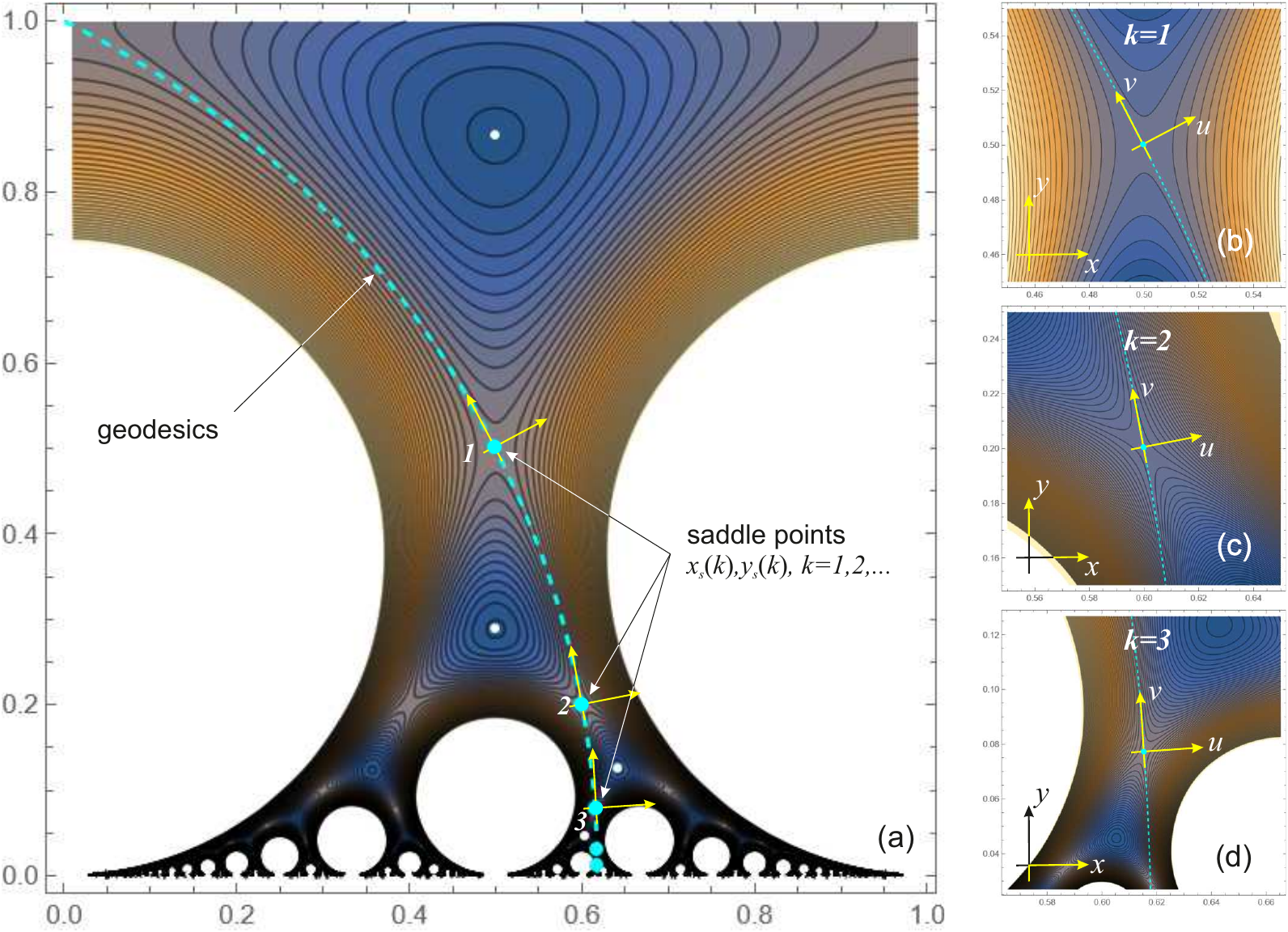}}
\caption{(a) Contour plot of the Riemann surface $U(x,y)=\ln \left(y^{1/4}|\eta(x+iy)|\right)$ in the region ($0.01<x<0.99$,\, $-0.35<y<-0.32$). White points are the same as in \fig{fig:03}, while cyan points designate the bifurcation points of the RG flow, and the dashed arc is the open geodesics passing through all saddles $(x_s(k),y_s(k))$ constituting the Fibonacci series ($k=0,...,\infty$) -- see \eq{eq:21}; (b)--(d) plots of the surface $U(x,y)$ in vicinity of three first saddles $(x_s(k),y_s(k))$ for $k=1,2,3$.}
\label{fig:04}
\end{figure}

The contour plot of $U(x,y)$ for $q=1$ in the region ($0.01<x<0.99$,\, $-0.35<y<-0.32$) is shown in \fig{fig:04}a. By $(x_s(k),y_s(k))$ we denote the coordinates of the saddle points ($k=1,2,...$), they are shown in cyan in \fig{fig:04} and have the following generic expression:
\be
x_s(k) = \frac{n_1 m_1 + n_2 m_2}{m_1^2 + m_2^2}; \qquad y_s = \frac{1}{m_1^2 + m_2^2}
\label{eq:20}
\ee
where $(m_1,m_2,n_1,n_2)$ are some integers. In particular, in \fig{fig:04}a we have depicted the following points constituting the Fibonacci sequence: $(x_s(1),y_s(1))=(\tfrac{1}{2}, \tfrac{1}{2})$, $(x_s(2),y_s(2))=(\tfrac{3}{5},\tfrac{1}{5})$, $(x_s(3),y_s(3))=(\tfrac{8}{13}, \tfrac{1}{13})$, $(x_s(4),y_s(4))=(\tfrac{21}{34}, \tfrac{1}{34})$, etc. From the topological point of view there is no difference between all these saddle points, however the orientation of saddles with respect to the $x$-axis is different and the geodesic (dashed cyan curve in \fig{fig:04}a), parameterized by the equation $y(x)=\sqrt{\frac{5}{4} - \left(x + \frac{1}{2}\right)^2}$, crosses saddle points with different $k$ at different angles. The coordinates $(x_s(k),y_s(k))$ of saddles which constitute the Fibonacci series are:
\be
\Big(x_s(k),y_s(k)\Big)= \left(\frac{G_1^{2k}-G_2^{2k}}{G_1^{2k+1}-G_2^{2k+1}}, \frac{\sqrt{5}}{G_1^{2k+1}-G_2^{2k+1}}\right); \qquad k=0,1,2,...,\infty
\label{eq:21}
\ee
where $G_1=\frac{1}{2}(1+\sqrt{5})$ and $G_2=\frac{1}{2}(1-\sqrt{5})$. 

To proceed, we expand the potential $U(x,y)$ in the vicinity of a saddle point $(x_s,y_s)$ and find the equation describing the corresponding surface $U(x,y)$ near $(x_s,y_s)\equiv (x_s(k),y_s(k)$ (to shorten forthcoming expressions we suppress the index $k$). The Taylor expansion of $U(x,y)$ up to the second order reads: 
\be
U(x-x_s, y-y_s) \approx U(x_s,y_s) + U_{xx}\,(x-x_s)^2+2U_{xy}\,(x-x_s)(y-y_s)+U_{yy}\,(y-y_s)^2
\label{eq:22}
\ee
where the derivatives $U_{xx}$, $U_{xy}=U_{yx}$, and $U_{yy}$ are taken at the point $(x_s,y_s)$. The first derivatives in the Taylor expansion \eq{eq:22} are nullified at the point $(x_s,y_s)$, and the condition $U_{xx}U_{yy} - U_{xy}^2<0$ ensures that the point $(x_s,y_s)$ is actually a saddle. Note also that the shift, $U(x_s,y_s)\equiv U(x_s(k),y_s(k))$, is one and the same constant for any $k$ and therefore can be suppressed. Define now the coefficients $U_{xx}=a_1$, $U_{x,y}=a_2$, $U_{yy}=a_3$ at every saddle point. These coefficients, $a_1=a_1(k)$, $a_2=a_2(k)$ and $a_3=a_3(k)$, depend on $k$, where $k=1,...,\infty$ in \eq{eq:21}. 

Derive now the RG flow in the complex $z=x+iy$ plane in the vicinity of saddle points $(x_s,y_s)$ of the surface $U(x,y)$ which plays the role of the $\beta$-function, as it has been shown in Section \ref{sect:02c}. Introducing the coordinates $u=x-x_s$ and $v=y-y_s$ and separating real and imaginary parts, we may write down the following first-order nonlinear differential equations describing the RG flow in the complex plane $w=u+iv$:
\be
\begin{cases}
\disp \frac{du}{d\ln \mu} = a_1 u^2 - a_3 v^2 \medskip \\ 
\disp \frac{dv}{d\ln \mu} = 2 a_2\, u v
\end{cases}
\label{eq:23}
\ee
where $\mu$ is the RG time. 

Equations \eq{eq:23} imply that the RG flow near the bifurcation points is fully determined by the topology of the $\beta$-function $U(x,y)$. It is worth mentioning that our construction is consistent with ideas expressed in works \cite{kaplan1,lutken,carpentier}. In particular, in \cite{kaplan1} the connection between the RG flows and the topological structure of the $\beta$-function has been discussed in the context of CFT/ADS$_2$ duality, while in \cite{lutken} and in \cite{carpentier} the equations for RG flows ideologically similar to \eq{eq:23} have been derived to describe the behavior of RG flows in the FQHE in the vicinity of critical points. The emergence of BKT fixed points in similar context has been also studied in \cite{fisher} for layered high-$T_c$ superconductors.

Before proceeding with the derivation of the solution of \eq{eq:23}, let us formulate the main idea behind our computations. Comparing the orientation of the $(u,v)$ coordinate system with respect to $(x,y)$ one for different $k$ (see \fig{fig:04}b,c,d, we can note that with increasing $k$ the $(u,v)$-system rotates in such a way that at $k\to \infty$ it coincides with the $(x,y)$ one. Computing explicitly the shape $U(x,y)$ in vicinity of the saddle point $(x_s(k), y_s(k))$ for $k=1,2,3$, we see that with $k\to\infty$ the coefficient $a_1=U_{xx}$ tends to zero, while the coefficients $a_2=U_{y}$ and $a_3=U_{yy}$ remain finite. To demonstrate this, we have depicted in \fig{fig:04}(b)-(d) the potentials $U(x,y)$ in vicinity of first three terms of the Fibonacci series for $k=1,2,3$:
\be
\begin{cases}
U_1(u,v) = -3.31 - {\bf 0.92} u^2 - 0.92 v^2 + 1.85 u v \medskip \\
U_2(u,v) = -3.31 + {\bf 2.60} u^2 + 23.39 v^2 - 15.58 u v \medskip \\
U_3(u,v) = -3.31 - {\bf 1.25} u^2 - 78.65 v^2 + 19.77 u v \medskip \\
\end{cases}
\label{eq:24}
\ee
One sees from \eq{eq:24} that with increasing $k$ the coefficient $a_1(k)$ in front of the term $u^2$ relatively decreases. Returning to \eq{eq:23} we see that when the coefficient $a_1$ is nullified, the corresponding equation coincides with the equation describing the RG flow in $XY$ model near the BKT transition point. Thus we expect that at $y\to 0$ which means strong compression of the fillotaxis lattice the corresponding lattice rearrangements (bifurcations) are closer and closer to the BKT transition.

Let us return now to \eq{eq:23} and provide its exact solution at any $a_1,a_2,a_3$. Dividing the first equation of \eq{eq:23} by the second one we can convert the system \eq{eq:23} into the following single equation
\be
\frac{du}{dv} = \frac{a_1}{2a_2}\frac{u}{v} - \frac{a_3}{2a_2}\frac{v}{u}
\label{eq:25}
\ee
Introducing the new function $h$ and writing $u = h v$, we arrive at the equation in which the variables can be separated:
\be
v\frac{dh}{dv} = \left(\frac{a_1}{2a_2}-1\right)h - \frac{a_3}{2a_2}h^{-1}
\label{eq:26}
\ee
Solving \eq{eq:25} we get
\be
\frac{a_2}{a_1 - 2 a_2} \ln\left(a_3 - (a_1 - 2 a_2) h^2\right) = \ln (G v)
\label{eq:27}
\ee
where $G$ remains invariant along the RG flow (i.e. $G$ does not depend on the scale $\mu$). Plugging the function $h=u/v$ in \eq{eq:27} and denoting $G^{a_1/a_2-2}$ by $\Delta$, we have
\be
a_3v^2-v^{a_1/a_2} \Delta = (a_1-2a_2)u^2
\label{eq:28}
\ee
Substituting $u(v)$ into the second equation in \eq{eq:23} and performing the integration, we obtain an non-explicit solution for $v(\mu)$
\begin{multline}
\frac{\sqrt{a_1-2 a_2}}{a_1 \sqrt{a_3 v^2-\Delta  v^{\frac{a_1}{a_2}}}} \Bigg(\left(a_1-2 a_2\right) a_3 v^2 \sqrt{1-\frac{\Delta  v^{\frac{a_1}{a_2}-2}}{a_3}} \, _2F_1\left(\frac{1}{2},\frac{a_2}{a_1-2 a_2};\frac{a_1-a_2}{a_1-2 a_2};\frac{v^{\frac{a_1}{a_2}-2} \Delta }{a_3}\right)+ \\ 2 a_2 \left(a_3 v^2-\Delta  v^{\frac{a_1}{a_2}}\Bigg)\right) = \ln \mu
\label{eq:29}
\end{multline}
Despite \eq{eq:29} looks rather complicated, it is essentially simplified in the limit of small $y$. Substituting $a_1=0$ into \eq{eq:26} we get the set of equations describing the RG flow in the $XY$-model in vicinity of the BKT transition. The critical scale (the correlation length) near the transition point is defined by the condition $\sqrt{-2a_2\Delta} \ln \mu_c \sim 1$ which implies the BKT-like dependence of the correlation length, $\mu_c$, on $\Delta$:
\be
\mu_c \sim e^{1/\sqrt{-2a_2 \Delta}}
\label{eq:30}
\ee
One can see from \eq{eq:24} that the coefficient $a_2$ in front of the $uv$ term periodically changes the sign. So, one can expect the signature of the BKT-like transition \eq{eq:30} when $a_2<0$. 

The physical meaning of encountered critical behavior could have the following interpretation. When the cylinder is squeezed along its principal axis, the corresponding lattice of repulsive particles experiences a set of successive  rearrangements (``bifurcations''). Each bifurcation is a collective effect that is accompanied by the melting of the lattice. Our analysis permits us to conjecture that these some of these bifurcations in the strong compression limit have signatures of Berezinsky-Kosterliz-Thouless (BKT) transtions. 

\section{Discussion}
\label{sect:04}

The main result of our work is as follows: we have provided arguments in support of the conjecture that bifurcations of the lattice formed by strongly repulsive particles equilibrated at the surface of a finite cylinder are the points of phase transitions in the thermodynamic limit. At strong cylinder compression equations describing RG flow near these transition points converge towards equations describing the Berezinskii-Kosterlitz-Thouless (BKT) behavior in the $XY$ model. The numerical check of this conjecture is highly demanded since our analysis is not restricted to the specific model of phyllotaxis. Relying on a general topology of the $\beta$-function in the vicinity of transition points one could expect the signature of BKT transition in the Fractional Quantum Hall Transitions (FQHT) at small disorder (small $\sigma_{xx}$). 

The main ingredient in our construction is the lattice potential $U(x,y)$ defined in \eq{eq:08} which possesses modular properties and is scale-invariant as it is shown in Section \ref{sect:02c}. The variables $x$ and $y$ are combined in one complex variable, $z=x+iy$, playing a role of a modular parameter. It has been mentioned in the Introduction that typically the real part, $x$, has a sense of a ``topological term'', while the imaginary part, $y$, deals with the contribution coming from the ``disorder''. Besides, the potential $U(x,y)$ has been constructed in a purely geometric way, and if the coordinate $x$ has a rather clear topological sense since it is related to the winding angle $\alpha$ according to \eq{eq:02}, the meaning of the coordinate $y$ as a disorder is far from obvious. To establish the connection of $y$ with the disorder we demonstrate that the function $\sqrt{-U(x,y)}$ has an interpretation as a spectral density in a well-known model of a spectral statistics of random walks on ensemble of intervals of length $n$ ($n=1,2,3,...)$, and $n$ is distributed exponentially with the density $Q_n(\beta) = (e^{\beta}-1)e^{-\beta n}$. It is shown below that the corresponding spectral density, $\rho(\lambda, \beta)$ coincides with the properly normalized value $W(x,y)=\sqrt{-U(x,y)}$ where the following change of variables is implied: $x = \tfrac{1}{\pi}\arccos\tfrac{\lambda-2}{2}$, $2\le \lambda \le 4$, and $y = g(\beta)$ (the function $g(\beta)$ will be discussed below). This relation allows to establish a clear-cut view on the link of $y$ with the disorder strength, $\beta$.

So, consider the spectral statistics of a discrete Laplace operator, $L$, on the interval of length $n$ with the periodic boundary conditions. Our desire is to compute the spectral density $\rho(\lambda)$ of $L$ on the ensemble of random intervals distributed with the probability density $Q_n(\beta)=(e^{\beta}-1)e^{-\beta n}$. The spectrum of the $n\times n$ periodic three-diagonal Laplacian matrix, $L_{n\times n}$ with entries 
[$a_{i,i}=2$, $a_{i,i+1}=a_{i+1,i}=1$ and $a_{i,j}=0$ otherwise ($\{i,j\}=1,...,n$)], reads
\be
\lambda_{j,n}=2-2\cos\frac{\pi j}{n+1} \quad (1\le j\le n)
\label{eq:32}
\ee
The spectral density $\rho(\lambda)$ of the ensemble of $n\times n$ periodic random matrices $L_{n\times n}$ can be written in a form of a resolvent:
\be
\rho(\lambda,\beta) = \lim_{n\to\infty}\frac{1}{n}\la \sum_{m=1}^n\sum_{j=1}^{m} \delta(\lambda-\lambda_{j,m}) \ra_{Q_m(\beta)} =  \lim_{\stackrel{n\to\infty}{\eps\to 0}} \frac{\eps}{\pi n} \sum_{m=1}^n Q_m(\beta) \sum_{j=1}^m \im\, \frac{1}{\lambda-\lambda_{j,m} - i\eps}
\label{eq:33}
\ee
where $\la ...\ra$ means averaging over the distribution $Q_m(\beta)=(e^{\beta}-1)e^{-\beta m}$, and the identity 
\be
\delta(x) = \frac{1}{\pi} \lim_{\eps\to+0} \im \frac{1}{x - i\eps}
\label{eq:34}
\ee
is used to regularize the $\delta$-function in \eq{eq:33}. Substituting \eq{eq:32} into \eq{eq:33}, we find the following expression for $\rho(\lambda)$:
\be
\rho(\lambda,\beta) = \lim_{\stackrel{n\to\infty}{\eps\to 0}} \frac{\eps(e^{\beta}-1)}{\pi n} \sum_{m=1}^{n} e^{-\beta m}
\sum_{j=1}^m\frac{1}{\left(\lambda-2+2\cos\frac{\pi j}{m+1}\right)^2+\eps^2}
\label{eq:35}
\ee
The function $\rho(\lambda, \beta)$ lies in the interval $[0\le \lambda \le 4]$, is symmetric and has maximum at the point $\lambda=2$. The spectral density $\rho(\lambda,\beta)$ in Eq.\eq{eq:35} matches at $0<\beta\ll 1$ the function $W(x,y)$ 
\be
W(x,y) = \left(\frac{12 y}{\pi}\right)^{1/2}\sqrt{-\ln \left(y^{1/4}|\eta(x+iy)|\right)}
\label{eq:36}
\ee
under the following change of variables in \eq{eq:36}: $x=\frac{1}{\pi} \arccos\frac{\lambda-2}{2}$, $y = g(\beta) \equiv h(n,\epsilon) \beta$, where $h(n,\eps)$ is some function of $n$ and $\eps$ (but not of $\beta$), and $0<\beta\ll 1$. To see this matching, it is convenient to compare the normalized functions, $\tilde{\rho}(\lambda,\beta)= \frac{\rho(\lambda,\beta)}{\rho(\lambda = 2,\beta)}$ and $\tilde{W}(\lambda,y) = \frac{W(\lambda,y)}{W(\lambda=2,y)}$. In \fig{fig:04}a,b we have plotted the normalized functions $\tilde{\rho}(\lambda,\beta)$ (panel (a)) with the following set of parameters: $n=200$, $\beta=7\times 10^{-2}$, and $n=200$, $\beta=7\times 10^{-2}$ (panel (b)) for $y=10^{-4}$. The values $\beta$ and $y$ are adjusted in such a way that both plots provide one and the same ``resolution cutoff'' (i.e. number of smallest peaks which can be still resolved).

\begin{figure}[ht]
\centerline{\includegraphics[width=16cm]{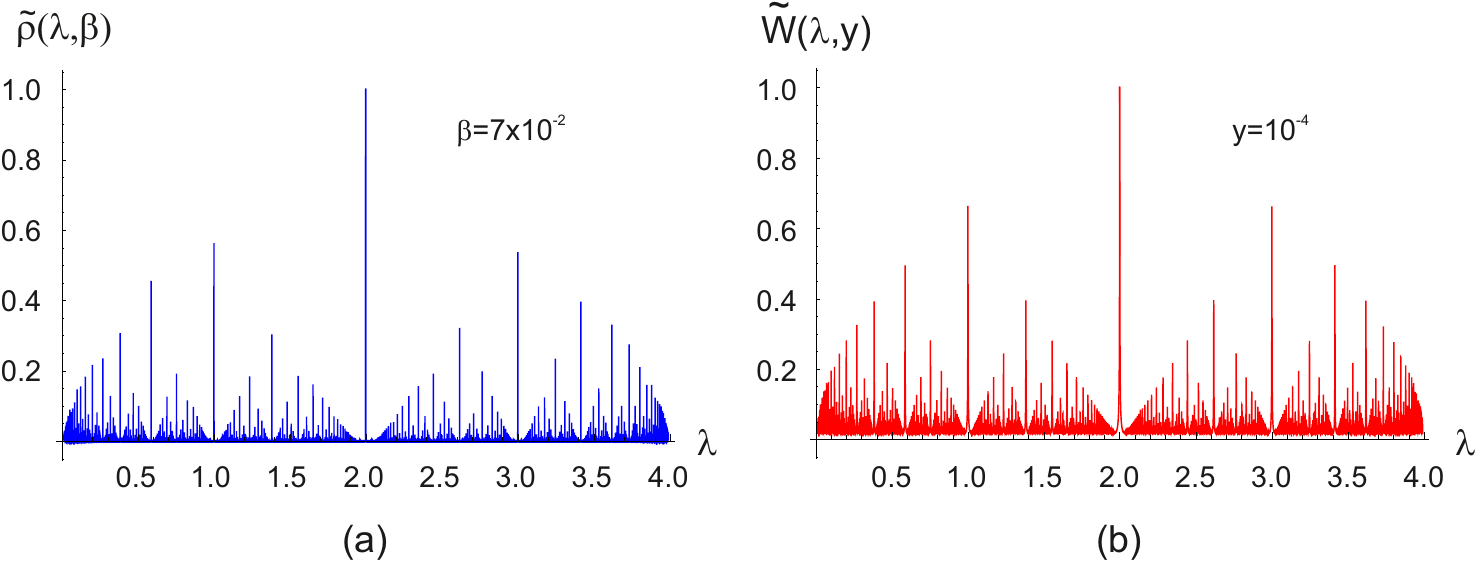}}
\caption{(a) Plot of the function $\tilde{\rho}(\lambda,\beta)$ for $n=200$, $\beta=7\times 10^{-2}$; (b) Plot of the function $n=200$, $\beta=7\times 10^{-2}$ for $y=10^{-4}$. The parameters $\beta$ and $y$ are adjusted to provide one and the same ``resolution cutoff''.}
\label{fig:05}
\end{figure}

The parameter $y$ in \eq{eq:36} has a sense of a ``resolution cutoff'' of the Dedekind relief. The relation between the strength of the disorder, $\beta$, and the cutoff, $y$, can be established using the following qualitative arguments. On one hand, the maximal denominator in \eq{eq:35}, $n$, defines the total number of peaks that can be resolved, $n_{max}$. The corresponding resolution cutoff can be estimated as $n_{max} \beta \sim 1$. On the other hand, the cutoff $y$ in \eq{eq:36} can be estimated as $y\sim 1/n_{max}$. Thus, in the limit $\beta\to 0$ one has a relation $y\approx \beta$. Thus, we see that the variable $y$ at $y\to 0$ can be interpreted as an effect of a disorder: as more pure the system ($\beta \to 0$), as more detailed fractal structure in \fig{fig:04} is be seen.

\begin{acknowledgments}
The authors are grateful to Alexander Gorsky for numerous fruitful discussions and highly acknowledge the hospitality of the staff of UWC Dilijan College for organizing an unforgettable meeting.
\end{acknowledgments}

\centerline{\rule{12cm}{0.4pt}}

\textsf{\small This text is a self-contained extract of a more general work ``Generalized Devil's staircase and RG flows'' (see \cite{agn}). The paper is written on the basis of the talk given by one of us (S.N.) in October 2022 at the workshop ``Statistical mechanics and its applications'' (Dilijan, Armenia).}

\bibliography{rg-bkt.bib}

\end{document}